# Perpendicular magnetic anisotropy tuning of macrospin-to-vortex transitions in Co-based artificial spin-vortex ice

Yu Maruyama,[1, 2] Amrit Kumar Mondal,[2] Bijaya Kharel,[2] Ryo Ohshima,[1,3] Jorge Puebla,[1,3] M. Benjamin Jungfleisch,[2] and Masashi Shiraishi[1, 3]

[1]*Department of Electronic Science and Engineering, Kyoto University, Kyoto 615-8510, Japan*

[2]*Department of Physics and Astronomy, University of Delaware, Newark, Delaware 19716, USA*

[3]*CSRN, Kyoto University, Kyoto 615-8510, Japan*

# Corresponding author: Yu Maruyama (maruyama.yuu.65w@st.kyoto-u.ac.jp), M. Benjamin Jungfleisch (mbj@udel.edu), and Masashi Shiraishi (shiraishi.masashi.4w@kyoto-u.ac.jp)

We investigate the macrospin-to-vortex (MS-to-V) transition in Co-based artificial spin-vortex ice (ASVI) in the presence of perpendicular magnetic anisotropy (PMA) by spin-wave spectroscopy. Detailed micromagnetic simulations using mumax3 reveal that the PMA modifies the magnetic energy landscape and facilitates vortex formation, suggesting that PMA can enhance the transition probability. To seek experimental validation of this hypothesis, we prepared Ti (3 nm)/Co (10 nm)/Ti (3 nm)/Pt (2 nm) (TCT) and Ti (3 nm)/Co (10 nm)/Pt (2 nm) (TCP) multilayer stacks. Vibrating sample magnetometry measurements confirm that the TCP film exhibits a larger PMA than the TCT film. Using these stacks, we then investigate the MS-to-V transition probability in ASVIs and found that TCP ASVIs exhibit a higher transition probability than TCT ASVIs, in agreement with the simulation prediction. These findings identify PMA as an effective design parameter for controlling vortex formation in ASVIs and provide a promising route toward task-dependent tuning of fading-memory properties for physical reservoir computing based on artificial spin lattices.

Artificial spin ice (ASI) provides an engineered nanomagnetic platform with a large number of metastable microstates, intrinsic physical memory, and microstate-dependent high-frequency dynamics.[1–4] These properties have recently motivated the use of ASI for unconventional computing, where magnetic memory and nonlinearities are used for information processing.[4–8] Among ASI-based systems, artificial spin-vortex ice (ASVI), which integrates both macrospin (MS) and vortex (V) states, has emerged as a promising platform for physical reservoir computing.[4] In the system, the probabilistic transition from the MS state to the V state gives rise to fading-memory characteristics, which is one of the key requirements for reservoir computing.[4,7–12]

Optimal memory capacity is essential for effective reservoir computing.[7–11] In ASVI, this capacity is governed by the MS-to-V transition probability $p$.[4] A larger $p$ shortens the memory of past inputs, whereas a smaller $p$ preserves the input history over more training cycles. Since the MS-to-V transition is governed by the relative stability of the MS and V states, a straightforward way to tune $p$ is to modify the nanomagnet geometry through shape anisotropy.[4] However, geometry-based tuning fixes the memory characteristics once the device is fabricated. Therefore, a material-based parameter that tunes $p$ without modifying the ASVI geometry would provide a flexible route toward task-adapting physical reservoir.[7]

One possibility to achieve this goal is to introduce perpendicular magnetic anisotropy (PMA) as a tuning parameter of $p$ by modifying the stability of the V states.[13,14] Since the V state contains an out-of-plane (OOP) magnetization component at the vortex core, PMA is expected to alter the energy balance between the MS and V states and thereby modulate $p$. Moreover, because interfacial PMA can be modulated by an electric field,[15] PMA engineering may provide a route to on-chip tuning of the memory capacity.

In this Letter, we demonstrate that PMA can indeed control the MS-to-V transition probability in ASVI. We show, using micromagnetic simulations and vector-network-analyzer-based ferromagnetic resonance (VNA-FMR) measurements, that increasing the PMA promotes vortex-state formation and enhances the probability. These results are further supported by VSM measurements evaluating the PMA and by magnetic force microscopy (MFM) measurements confirming the MS-to-V transition.

To clarify whether PMA promotes the MS-to-V transition, we performed micromagnetic simulations using MuMax3.[16,17] We evaluated the relative stability of the V and MS states using the energy difference[4]

$$\Delta E_{\mathrm{V-MS}} = E_{\mathrm{total}}(\mathrm{MS}) - E_{\mathrm{total}}(\mathrm{V}), \tag{1}$$

where $E_{total} = E_{demag} + E_{exch} + E_{anis}$ is the total energy ($E_{demag}$, $E_{exch}$ and $E_{anis}$ denote the demagnetization, exchange, and magnetic anisotropy energies, respectively). A more positive (negative) $\Delta E_{V\text{-}MS}$ indicates that the V (MS) state is energetically more favorable. The simulations were carried out for a single stadium-shaped nanomagnet with dimensions of 600 nm × 220 nm × 10 nm, using a saturation magnetization $M_{sat}$ = 1.6 MA/m and an exchange stiffness $A_{ex}$ = 20 pJ/m.[18] Figure 1 (a) shows the $K_u$ dependence of $\Delta E_{V\text{-}MS}$. $\Delta E_{V\text{-}MS}$ increases monotonically with increasing $K_u$, indicating that the introduction of $K_u$ facilitates the formation of the V state. Figure 1(b), on the other hand, demonstrates the $K_u$ dependence of $E_{total}$ for the V and MS states. Although the $E_{total}$ decreases for both states, the decrease is more pronounced for the V state. In Fig. 1(c), we highlight the contributions from $E_{demag}$, $E_{exch}$ and $E_{anis}$. These energy components are more sensitive to $K_u$ in the V state than in the MS state as can be seen by comparing the left with the right panel of Fig. 1(c). The observed behavior can be attributed to the OOP magnetization component localized at the vortex core.[13,14] Since PMA energetically favors OOP magnetization, the V state benefits more from increasing $K_u$. This tendency is consistent with previous studies on magnetic nanocylinders, where PMA was reported to expand the vortex-stable region in the phase diagram[13] and increase the vortex-core size.[14] Our findings suggest that a similar influence of PMA can also occur in the stadium-shaped nanomagnets in ASVI, where a moderate increase in $K_u$ promotes the vortex-state formation and thereby enhances $p$.

Before experimentally realizing ASVI with different PMA strengths, we studied the magnetic properties of two Co-based multilayer thin-film stacks in which the upper Co interfaced with either Pt or Ti/Pt while the bottom interface was kept identical. We fabricated Ti (3 nm)/Co (10 nm)/Ti (3 nm)/Pt (2 nm) (TCT) and Ti (3 nm)/Co (10 nm)/Pt (2 nm) (TCP) stacks using electron beam deposition on sapphire substrates. The deposition pressure was $10^{-6}$ Pa, and the deposition rates were 0.02 nm/s for Co and Ti and 0.01 nm/s for Pt. The Co/Pt interface was chosen to enhance PMA because the interfacial anisotropy energy $K_s$ at the Co/Pt interface has been reported to be $K_s^{Co/Pt} = 0.6 – 1.4$ mJ/m$^2$,[19,20] which is noticeably larger than that at the Co/Ti interface, $K_s^{Co/Ti}$ = 0.23 mJ/m$^2$.[21] Thus, TCP is expected to have a larger $K_u$ than TCT. The same Ti underlayer was used as a seed layer for both sample stacks to avoid underlayer induced variations in the magnetic properties of Co.[22,23] In particular, we chose Ti because it enables low-damping ferromagnetic resonance (FMR) in thin Co film.[22]

To experimentally confirm the difference in $K_u$ between the TCT and TCP stacks, the magnetic anisotropies of the continuous films were characterized by VSM. Figures 2(a) shows the VSM results for OOP and in-plane (IP) magnetic fields, respectively. The linear background, estimated from the saturated region ($0.5 < |\mu_0 H| < 2.5$ T for IP and $1.8 < |\mu_0 H| < 2.5$ T for OOP, where $\mu_0$ is the vacuum permeability and $H$ is the applied magnetic field), was subtracted. Both films exhibit IP easy axis behavior. For thin films with IP easy axis, $K_u$ is expressed as $K_u = K_{eff} + \mu_0 M_s^2/2$, where $M_s$ is the saturation magnetization, and $K_{eff}$ is the effective anisotropy energy density. $K_{eff}$ was evaluated from the area between the OOP and IP magnetization curves, with a negative sign.[23] The values obtained from the positive- and negative-field branches were averaged, giving $K_{eff} = (-0.678 \pm 0.003)M_s$ and $(-0.634 \pm 0.002)M_s$ J/m$^3$ for the TCT and TCP films, respectively. Using $M_s = 1611 \pm 2$ kA/m, estimated from the IP magnetization curves, $K_u$ was determined to be $539 \pm 6$ and $609 \pm 5$ kJ/m$^3$ for the TCT and TCP films, respectively. Assuming that the observed difference in $K_u$ mainly originates from the difference in the upper interface, the difference $\Delta K_u$ for Co thickness of $t_{Co} = 10$ nm corresponds to an interfacial anisotropy difference of $\Delta K_s = K_s^{Co/Pt} - K_s^{Co/Ti} \sim \Delta K_u t_{Co} \sim 0.70 \pm 0.08$ mJ/m$^2$, which is consistent with the difference expected from the reported $K_s$ values discussed above.[19–21] Our VSM results confirm that TCP stack has a larger $K_u$ than TCT stack and, hence, is a promising choice for investigating the effect of PMA on the MS-to-V transition.

We then fabricated ASVI devices using electron-beam lithography, electron-beam deposition, and a lift-off technique. Figure 2(b) shows the schematic illustration of the ASVI device. Stadium-shaped nanomagnets composed of either TCT or TCP were arranged in a 45°-tilted square-lattice geometry directly on a coplanar waveguide (CPW).[24] The CPW, made of Ti (3 nm)/Au (100 nm), was fabricated on a thermally oxidized Si substrate. During both the training process and the VNA-FMR measurements, the external magnetic field $B_{ext}$ was applied along the long side of the CPW.

Figure 2(c) shows the design parameters of the nanomagnets. The widths of the stadium-shaped nanomagnets were varied as 220, 230, and 240 nm, while the length and the edge-to-edge distance between neighboring nanomagnets were fixed at 600 and 80 nm, respectively. Wider nanomagnets are expected to form the V state more easily because of their reduced shape anisotropy.[4] Figure 2(d) shows an exemplary scanning electron microscopy (SEM) image of the 220 nm-wide TCT ASVI. The SEM image confirms that well-defined stadium-shaped nanomagnets were successfully fabricated.

The MS-to-V transition occurs probabilistically during magnetization switching of the MS state.[4] To introduce the V state, the ASVI was first saturated by applying $B_{ext} = 180$ mT. Subsequently, a minor field loop between $-B_{train}$ and $+B_{train}$ was applied $N$ times as the training process. $B_{train}$ was chosen to be in between the MS switching field and the vortex annihilation field. In this study, we set $B_{train}$ between 20 and 22 mT depending on the ASVIs. A detailed discussion of the selection of $B_{train}$ is provided in the Supplementary Material.

In Fig. 3(a), a series of MFM images of the TCT ASVI (nanomagnet width: 220 nm) are shown for an increasing number of training cycles $N$. For $N = 0$, all nanomagnet MS were aligned along $B_{ext}$, forming a type-II ASI configuration. As $N$ increased, the population of the MS state gradually decreased, and approximately 86% of the nanomagnets transitioned to the V state after $N = 10$ loops. This gradual transition can be interpreted as a fading-memory behavior in the present 10-nm-thick Co-based ASVI, consistent with previous observations in 20-nm-thick Py-based ASVI,[4] although the transition probability differs between the two systems.

Because the MFM observation area (12 µm × 12 µm) covers only a small fraction of the more than 82,000 nanomagnets patterned on the CPW signal line (20 µm × 1500 µm), VNA-FMR was used to probe the collective behavior of an extended network and thereby obtain statistically significant information. Figure 3(b) shows the FMR spectra defined as $\Delta S_{21}(B_{ext}) = S_{21}(B_{ext}) - S_{21}(180\ \mathrm{mT})$, where $S_{21}(B_{ext})$ is the transmission parameter measured under $B_{ext}$. The spectra were acquired while sweeping $B_{ext}$ from +10 to −10mT. For $N = 0$, the spectrum shows two modes characteristic of a conventional ASI: a higher-frequency bulk-like mode and a lower-frequency edge mode.[4,24] With increasing $N$, the spectral profile evolves with the transition from the MS state to the V state, as confirmed by MFM. At $N = 10$, where most nanomagnets are expected to be in the V state, a characteristic inverted-V-shaped spectral feature was observed in the 4–6 GHz frequency range. In the V state, bulk-like modes can be localized on both sides of the vortex core, where the local magnetization directions are opposite. When $B_{ext}$ is applied, the vortex core is displaced from the center, causing one localized mode region to expand while the other shrink depending on the sign of $B_{ext}$. These bulk-like modes can appear at lower frequencies than that of the MS state and form an X-shaped spectral feature.[4] In our measurements, the bulk-like mode associated with the expanded mode region is likely to dominate, making the lower half of the X-shaped spectral feature appear as an inverted-V-shaped feature. These results indicate that the evolution of the VNA-FMR spectra is linked to the gradual MS-to-V transition.

In the following, we will analyze this transition behavior quantitatively. Figure 3(c) shows the $N$ dependence of $\Delta S_{21}$ at $B_{ext} = 0$ mT for different $N$. The spectra exhibit two peaks, which we fit using a two-mode Lorentzian function with symmetric and asymmetric components,

$$\Delta S_{21}(f) = \sum_{i=1}^{2} L_i(f) + b, \tag{2}$$

$$L_i(f) = \frac{L_{\mathrm{S}i}\Delta_i^2 + L_{\mathrm{A}i}\Delta_i(f - f_i)}{(f - f_i)^2 + \Delta_i^2}. \tag{3}$$

Here, $L_{Si}$ and $L_{Ai}$ are the symmetric and antisymmetric components, respectively, $f_i$ is the resonance frequency, $\Delta_i$ is the half width at half maximum, and $b$ is an offset corresponding to the background signal. The indices $i = 1$ and $i = 2$ denote the lower- and higher-frequency modes, respectively. The solid black lines in Fig. 3(c) represent the fitting results. This two-mode model reproduces the measured spectra reasonably well.

Figure 3(d) shows the $N$ dependence of $L_{S2}$, which shows the most pronounced change in the MS-to-V transition. The intensity of $L_{S2}$ gradually decreases with increasing $N$, reflecting the reduction of the MS-state population during the training process. Assuming a constant $p$ for each training loop, the $N$ dependence of $L_{S2}$ is fitted using

$$L_{S2}(N) = (L_0 - L_\infty)(1 - p)^N + L_\infty, \tag{4}$$

where $L_0$ is the initial amplitude associated with the MS state and $L_\infty$ is the offset contribution from the V state and the residual MS state. The solid line in Fig. 3(d) represents the fitting results, from which $p$ was extracted.

In Fig. 4(a), we extract the $N$ dependence of the normalized $L_{S2}$ for both TCT and TCP ASVIs. The intensity of $L_{S2}$ gradually decreases with increasing $N$ for all widths, indicating a continuous MS-to-V conversion during training process. Figure 4(b) summarizes the extracted $p$ as a function of nanomagnet width. For both stacks, $p$ increases with increasing width. This trend is consistent with the previous studies showing that $\Delta E_{V\text{-}MS}$ increases with increasing nanomagnet width.[4] The enhancement of $\Delta E_{V\text{-}MS}$ in wider nanomagnets attributed to the reduced shape anisotropy, which facilitates vortex-state formation.[4] More importantly, TCP ASVIs exhibit a larger $p$ than TCT ASVIs for all widths. This result is consistent with the trend of $\Delta E_{V\text{-}MS}$ shown in Fig. 1(a), supporting the interpretation that enhanced PMA increases $p$. Thus, $p$ can be controlled by both nanomagnet geometry and $K_u$.

In summary, we investigated the PMA dependence of the MS-to-V transition probability in TCT and TCP ASVI and its effect on the spin dynamics. Micromagnetic simulations showed that a moderate $K_u$ modifies the energy landscape by reducing the energy cost of a vortex-state formation, favoring the MS-to-V transition. Experimentally, the hypothesis was supported by three complementary measurements. VSM measurements confirmed that the TCP stack has a larger $K_u$ than the TCT stack, verifying that the stack design successfully tunes the PMA. MFM measurements then visualized the formation of the vortex state after training process. Finally, VNA-FMR measurements revealed that FMR spectra are linked to the magnetic state of the ASVI, and the estimated $p$ was higher in the TCP ASVI. Taken together, our combined experimental and simulation results demonstrate that the fading-memory characteristics of ASVI can be tuned by not only geometry but also PMA. Because $p$ controls the memory capacity relevant to reservoir computing, PMA engineering offers a materials-based route toward task-adapting ASVI reservoir devices. The electric-field tunability of interfacial PMA[15] provides an additional and suitable pathway toward on-chip optimization of ASI-based reservoir computing.

The Supplementary Material provides the MFM images of the 220-nm-wide TCP ASVI, the FMR spectra of all samples, and detailed discussion of the selection of $B_{\mathrm{train}}$.

## ACKNOWLEDGMENTS

This work was supported by JST SPRING (Grant No. JPMJSP2110), MEXT Initiative to Establish Next-generation Novel Integrated Circuits Centers (X-NICS) (Grant No. JPJ011438), and the Spintronics Research Network of Japan (Spin-RNJ). Work at Delaware including micromagnetic simulations, VSM, MFM, and VNA-FMR measurements were supported by the National Science Foundation through the University of Delaware Materials Research Science and Engineering Center (DMR-2011824). AKM and BK were supported by the National Science Foundation under Grant No. 2339475. MBJ acknowledges the JSPS Invitational Fellowship for Researcher in Japan.

## AUTHOR DECLARATIONS

### Conflict of Interest

The authors have no conflicts to disclose.

### Author Contributions

**Yu Maruyama**: Conceptualization (equal); Formal analysis (equal); Investigation (equal); Methodology (equal); Software (equal); Validation (equal); Visualization (equal); Writing – original draft (equal). **Amrit Kumar Mondal**: Formal analysis (equal); Investigation (equal); Visualization (equal); Writing – review & editing (equal). **Bijaya Kharel**: Methodology (equal); Investigation (equal); Writing – review & editing (equal). **Ryo Ohshima**: Writing – review & editing (equal). **Jorge Puebla**: Writing – review & editing (equal). M. **Benjamin Jungfleisch**: Conceptualization (equal); Methodology (equal); Validation (equal); Supervision (equal); Writing – review & editing (equal). **Masashi Shiraishi**: Supervision (equal); Writing – review & editing (equal).

## DATA AVAILABILITY

The data that support the findings of this study are available from the corresponding authors upon reasonable request.

## FIGURES

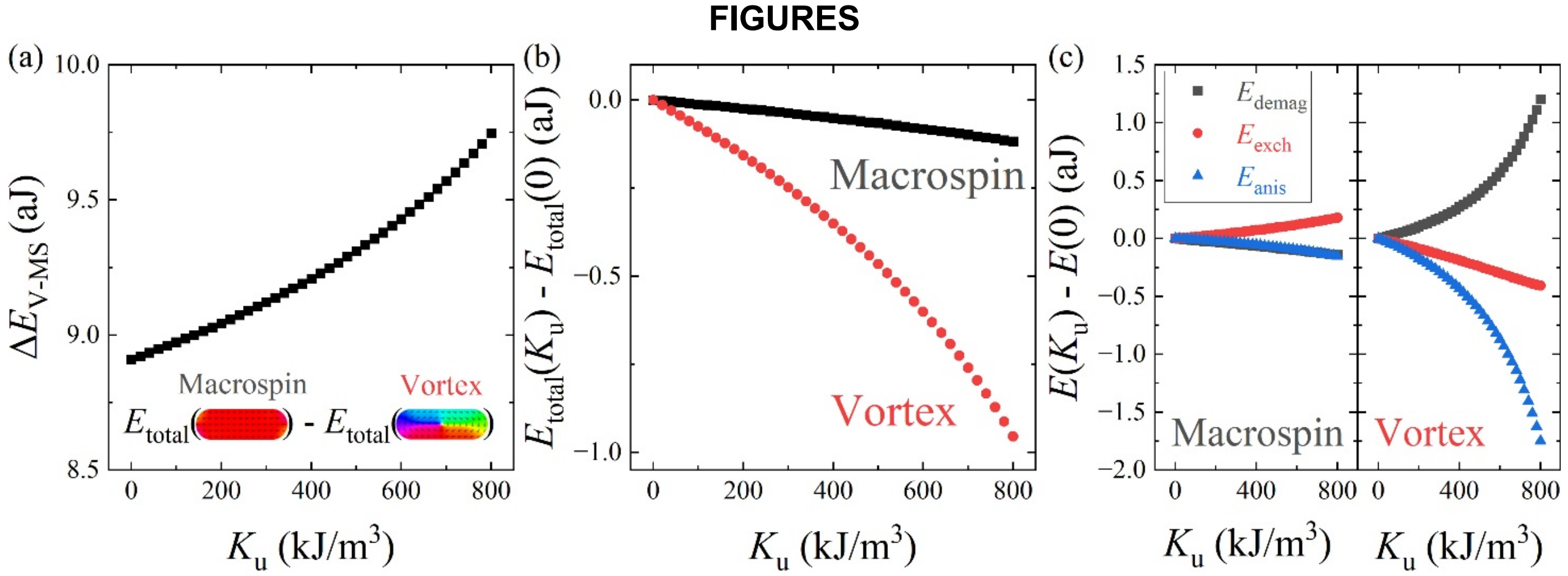

FIG. 1. $K_u$ dependence of (a) $\Delta E_{V\text{-}MS}$, (b) $E_{total}$ for the MS state (black squares) and V state (red circles), and (c) $E_{demag}$ (black squares), $E_{exch}$ (red circles), and $E_{anis}$ (blue triangles) for the MS and V states, obtained from MuMax3 simulation. These results indicate that the energy of the V state responds more sensitively to $K_u$ than that of the MS state, resulting in a larger $\Delta E_{V\text{-}MS}$ for moderate $K_u$.

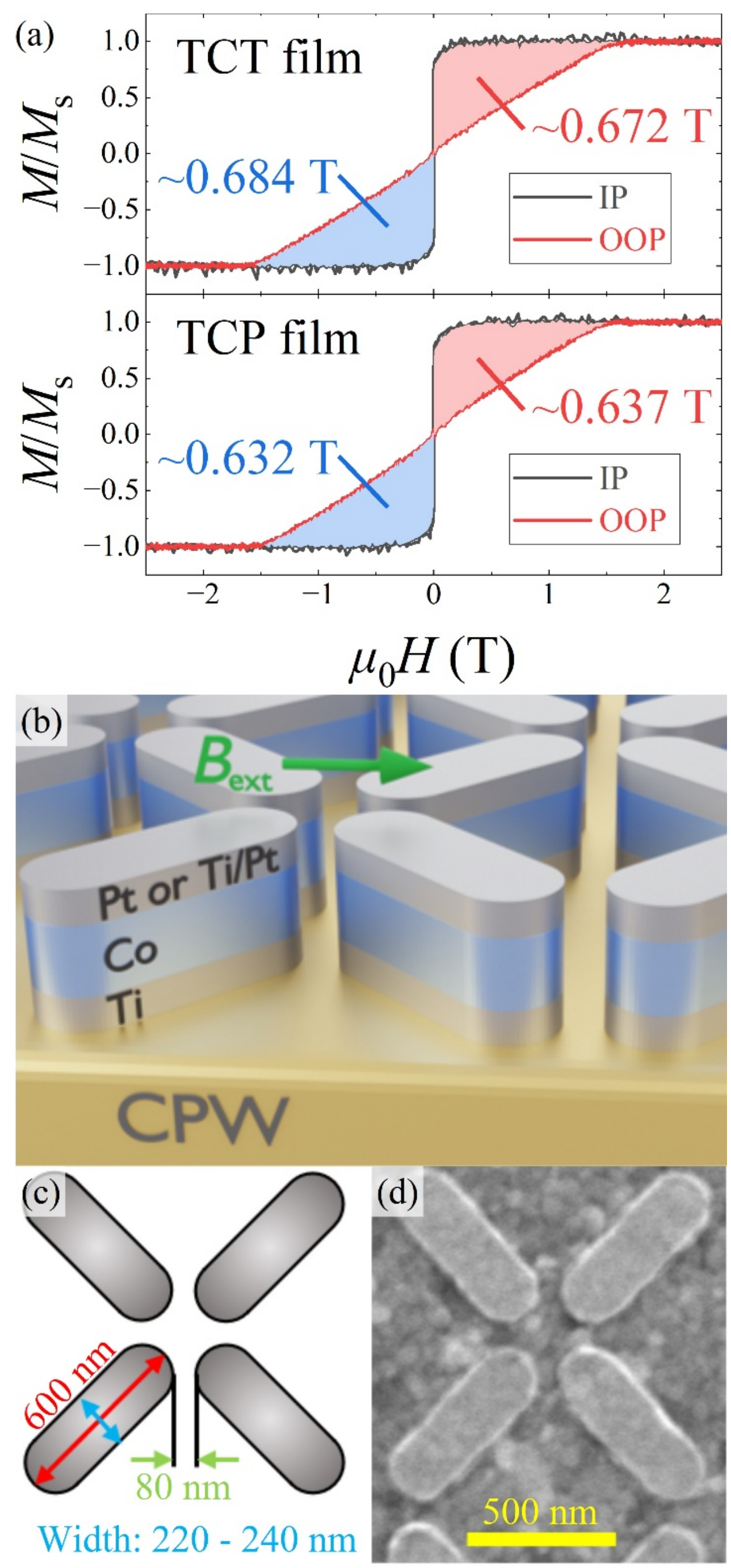

FIG. 2. (a) IP and OOP VSM results for the TCT and TCP films. Colored areas were used to determine the $K_{eff.}$ (b) Schematic illustration of the ASVI. The ASI is patterned on the signal line. The external field $\boldsymbol{B}_{ext}$ is applied along the long side of the CPW. (c) Schematic illustration of the ASI design. Stadium-shaped nanomagnets with different width were prepared. (d) Corresponding SEM image of the 220-nm-width ASVI.

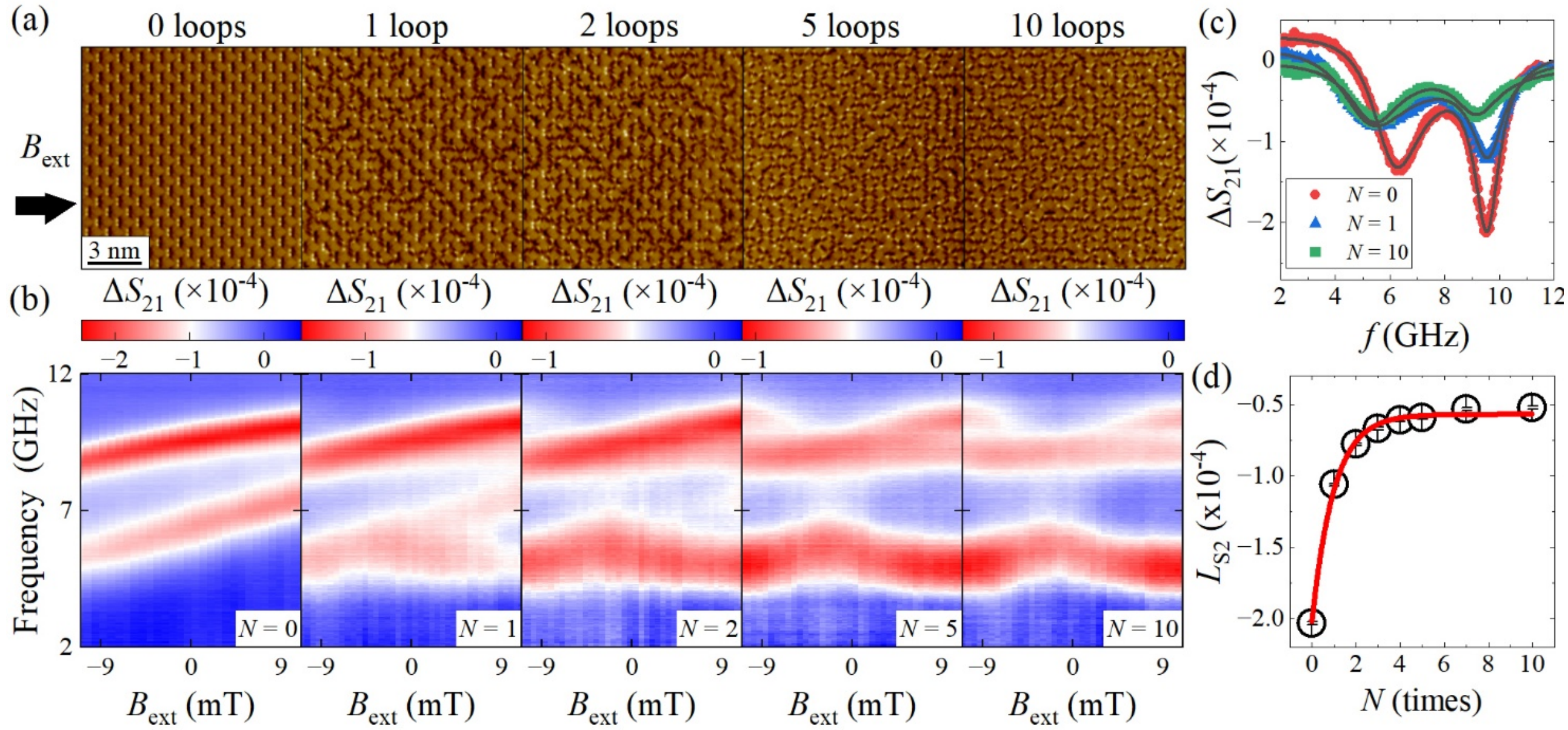

FIG. 3. (a) 12 μm ×12 μm MFM images and (b) differential FMR heatmaps of the 600 nm × 220 nm TCT ASVI obtained after $N$ minor loops ($N$ = 0, 1, 2, 5 and 10). The MFM images show that the vortex-state population increases with increasing $N$. Correspondingly, the FMR spectra gradually evolve from the macrospin-dominated response to the vortex-dominated response. (c) Differential FMR spectra of the 600 nm × 220 nm TCT ASVI measured at $B_{ext}$ = 0 mT after the training process. Black lines show the fitting results using Eq. (2) in the main text. (d) $N$ dependence of $L_{S2}$. red line shows the fitting results using Eq. (4) in the main text. The monotonic decrease in $L_{S2}$ reflects the progressive MS-to-V transition during training process.

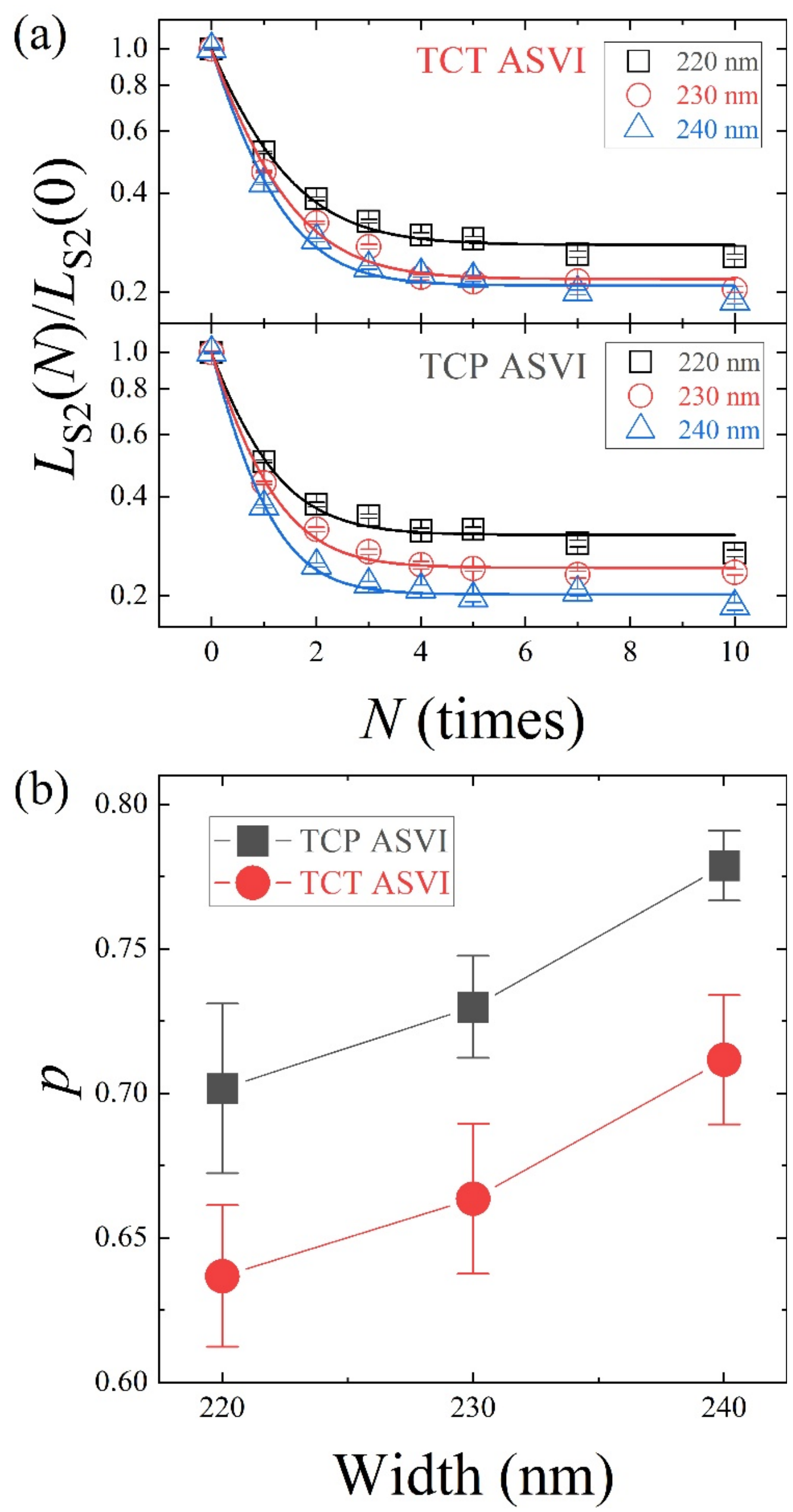

FIG. 4. (a) $N$ dependence of Normalized $L_{S2}$ of the TCT and TCP ASVIs with different width of 220 (black squares) 230 (red circles) and 240 (blue triangles) nm. Solid lines show the fitting results using Eq. (4). (b) Width dependence of $p$ of the TCT and TCP ASVIs. The increase in $p$ with both nanomagnet width and $K_u$ demonstrates that the fading-memory characteristics of ASVI can be tuned by both geometry and PMA.